\documentclass[twocolumn]{aastex7}

\usepackage{graphicx}
\usepackage{amsmath}
\usepackage{hyperref}
\usepackage{longtable}
\usepackage{booktabs}

\newcommand{\te}{t_{\rm E}}
\newcommand{\thetae}{\theta_{\rm E}}

\newcommand{\pie}{\pi_{\rm E}}

\newcommand{\pien}{\pi_{{\rm E},N}}
\newcommand{\piee}{\pi_{{\rm E},E}}
\newcommand{\dl}{D_{\rm L}}
\newcommand{\ds}{D_{\rm S}}
\def\e{{\rm E}}

\definecolor{brown}{rgb}{0.59, 0.29, 0.0}
\definecolor{darkgreen}{rgb}{0.0, 0.42, 0.24}
\definecolor{darkblue}{rgb}{0.01, 0.31, 0.59}
\definecolor{darkblue}{rgb}{0.0, 0.25, 0.42}
\definecolor{blue}{rgb}{0.0,0.0,1.0}
\definecolor{green}{rgb}{0.0,1.0,0.0}

\begin{document}

\title{KMT-2026-BLG-0083L: A Two-Jovian-Planet System Orbiting an M Dwarf Discovered by Microlensing}
\shorttitle{KMT-2026-BLG-0083}


\author{Cheongho Han}
\altaffiliation{KMTNet Collaboration}
\affiliation{Department of Physics, Chungbuk National University, Cheongju 28644, Republic of Korea}
\email{cheongho@astroph.chungbuk.ac.kr}
\author{Chung-Uk Lee}
\altaffiliation{KMTNet Collaboration}
\affiliation{Korea Astronomy and Space Science Institute, Daejon 34055, Republic of Korea}
\email{leecu@kasi.re.kr}
\author[0000-0002-1287-6064]{Zhixing Li}
\altaffiliation{DREAMS Collaboration}
\affiliation{Department of Astronomy, Westlake University, Hangzhou 310030, Zhejiang Province, China}
\email{lizhixing@westlake.edu.cn}
\author{Weicheng Zang}
\altaffiliation{KMTNet Collaboration}
\altaffiliation{DREAMS Collaboration}
\affiliation{Department of Astronomy, Westlake University, Hangzhou 310030, Zhejiang Province, China}
\email{zangweicheng@westlake.edu.cn}
\author{Andrzej Udalski} 
\altaffiliation{OGLE Collaboration}
\affiliation{Astronomical Observatory, University of Warsaw, Al.~Ujazdowskie 4, 00-478 Warszawa, Poland}
\email{udalski@astrouw.edu.pl} 
\author{Ian A. Bond}
\altaffiliation{PRIME Collaboration}
\affiliation{School of Mathematical and Computational Sciences, Massey University, Auckland 0745, New Zealand}
\email{i.a.bond@massey.ac.nz}
\author{Yoon-Hyun Ryu}
\altaffiliation{KMTNet Collaboration}
\affiliation{Korea Astronomy and Space Science Institute, Daejon 34055, Republic of Korea}
\email{yhryu@kasi.re.kr}
\author[0000-0002-0856-3663]{Steve Heathcote}
\altaffiliation{DREAMS Collaboration}
\affiliation{Cerro Tololo Inter-American Observatory/NSF NOIRLab, Casilla 603, La Serena, Chile}
\email{steve.heathcote@noirlab.edu}
\collaboration{14}{(Leading authors)}
\author{Michael D. Albrow}   
\affiliation{University of Canterbury, Department of Physics and Astronomy, Private Bag 4800, Christchurch 8020, New Zealand}
\email{michael.albrow@canterbury.ac.nz}
\author{Sun-Ju Chung}
\affiliation{Korea Astronomy and Space Science Institute, Daejon 34055, Republic of Korea}
\email{sjchung@kasi.re.kr}
\author{Andrew Gould}
\affiliation{Department of Astronomy, Ohio State University, 140 West 18th Ave., Columbus, OH 43210, USA}
\email{gould.34@osu.edu}
\author{Youn Kil Jung}
\affiliation{Korea Astronomy and Space Science Institute, Daejon 34055, Republic of Korea}
\affiliation{University of Science and Technology, Daejeon 34113, Republic of Korea}
\email{younkil21@gmail.com}
\author{Kyu-Ha Hwang}
\affiliation{Korea Astronomy and Space Science Institute, Daejon 34055, Republic of Korea}
\email{kyuha@kasi.re.kr}
\author{Hongjing Yang}
\affiliation{Department of Astronomy, Westlake University, Hangzhou 310030, Zhejiang Province, China}
\email{yanghongjing@westlake.edu.cn}
\author{Yossi Shvartzvald}
\affiliation{Department of Particle Physics and Astrophysics, Weizmann Institute of Science, Rehovot 76100, Israel}
\email{yossishv@gmail.com}
\author{In-Gu Shin}
\affiliation{Department of Astronomy, Westlake University, Hangzhou 310030, Zhejiang Province, China}
\email{ingushin@gmail.com}
\author{Doeon Kim}
\affiliation{Department of Physics, Chungbuk National University, Cheongju 28644, Republic of Korea}
\email{qso21@hanmail.net}
\author{Dong-Jin Kim}
\affiliation{Korea Astronomy and Space Science Institute, Daejon 34055, Republic of Korea}
\email{keaton03@kasi.re.kr}
\author{Byeong-Gon Park}
\affiliation{Korea Astronomy and Space Science Institute, Daejon 34055, Republic of Korea}
\email{bgpark@kasi.re.kr}
\author{Richard W. Pogge}
\affiliation{Department of Astronomy, Ohio State University, 140 West 18th Ave., Columbus, OH 43210, USA}
\email{pogge.1@osu.edu}
\collaboration{20}{(KMTNet Collaboration)}
\author[0000-0003-4625-8595]{Qiyue Qian}
\affiliation{Department of Astronomy, Westlake University, Hangzhou 310030, Zhejiang Province, China}
\affiliation{Department of Astronomy, Tsinghua University, Beijing 100084, China}
\email{qqy22@mails.tsinghua.edu.cn}
\author[0009-0003-7681-3702]{Yaosong Yu}
\affiliation{Department of Astronomy, Westlake University, Hangzhou 310030, Zhejiang Province, China}
\email{ayuyaosong@gmail.com}
\author[0000-0001-5651-9440]{Yuchen Tang}
\affiliation{Department of Astronomy, Westlake University, Hangzhou 310030, Zhejiang Province, China}
\email{tangyuchen@westlake.edu.cn}
\author[0009-0005-0410-8451]{Yuxin Shang}
\affiliation{Department of Astronomy, Tsinghua University, Beijing 100084, China}
\email{shangyx22@mails.tsinghua.edu.cn}
\author{Tomas Ahumada}
\affiliation{Cerro Tololo Inter-American Observatory/NSF NOIRLab, Casilla 603, La Serena, Chile}
\email{tomas.ahumada@noirlab.edu}
\author[0000-0003-1587-3931]{imothy Abbott}
\affiliation{Cerro Tololo Inter-American Observatory/NSF NOIRLab, Casilla 603, La Serena, Chile}
\email{tim.abbott@noirlab.edu}
\author[0000-0002-2651-7038]{Guillermo Damke}
\affiliation{Cerro Tololo Inter-American Observatory/NSF NOIRLab, Casilla 603, La Serena, Chile}
\email{guillermo.damke@noirlab.edu}
\author[0000-0003-4432-5037]{Konstantina Boutsia}
\affiliation{Cerro Tololo Inter-American Observatory/NSF NOIRLab, Casilla 603, La Serena, Chile}
\email{konstantina.boutsia@noirlab.edu}
\author[0000-0001-6455-9135]{Alfredo Zenteno}
\affiliation{Cerro Tololo Inter-American Observatory/NSF NOIRLab, Casilla 603, La Serena, Chile}
\email{alfredo.zenteno@noirlab.edu}
\author[0000-0001-8317-2788]{Shude Mao}
\affiliation{Department of Astronomy, Westlake University, Hangzhou 310030, Zhejiang Province, China}
\email{shude.mao@westlake.edu.cn}
\author[0009-0007-0032-4098]{Andong Xu}
\affiliation{Department of Astronomy, Westlake University, Hangzhou 310030, Zhejiang Province, China}
\email{xuandong@westlake.edu.cn}
\collaboration{100}{(DREAMS Collaboration)}
\author{Przemek Mr{\'o}z}
\affiliation{Astronomical Observatory, University of Warsaw, Al.~Ujazdowskie 4, 00-478 Warszawa, Poland}
\email{pmroz@astrouw.edu.pl}
\author{Micha{\l} K. Szyma{\'n}ski}
\affiliation{Astronomical Observatory, University of Warsaw, Al.~Ujazdowskie 4, 00-478 Warszawa, Poland}
\email{msz@astrouw.edu.pl}
\author{Jan Skowron}
\affiliation{Astronomical Observatory, University of Warsaw, Al.~Ujazdowskie 4, 00-478 Warszawa, Poland}
\email{jskowron@astrouw.edu.pl}
\author{Rados{\l}aw Poleski} 
\affiliation{Astronomical Observatory, University of Warsaw, Al.~Ujazdowskie 4, 00-478 Warszawa, Poland}
\email{radek.poleski@gmail.co}
\author{Igor Soszy{\'n}ski}
\affiliation{Astronomical Observatory, University of Warsaw, Al.~Ujazdowskie 4, 00-478 Warszawa, Poland}
\email{soszynsk@astrouw.edu.pl}
\author{Pawe{\l} Pietrukowicz}
\affiliation{Astronomical Observatory, University of Warsaw, Al.~Ujazdowskie 4, 00-478 Warszawa, Poland}
\email{pietruk@astrouw.edu.pl}
\author{Szymon Koz{\l}owski} 
\affiliation{Astronomical Observatory, University of Warsaw, Al.~Ujazdowskie 4, 00-478 Warszawa, Poland}
\email{simkoz@astrouw.edu.pl}
\author{Krzysztof A. Rybicki}
\affiliation{Astronomical Observatory, University of Warsaw, Al.~Ujazdowskie 4, 00-478 Warszawa, Poland}
\email{krybicki@astrouw.edu.pl}
\author{Patryk Iwanek}
\affiliation{Astronomical Observatory, University of Warsaw, Al.~Ujazdowskie 4, 00-478 Warszawa, Poland}
\email{piwanek@astrouw.edu.pl}
\author{Krzysztof Ulaczyk}
\affiliation{Department of Physics, University of Warwick, Gibbet Hill Road, Coventry, CV4 7AL, UK}
\email{kulaczyk@astrouw.edu.pl}
\author{Marcin Wrona}
\affiliation{Astronomical Observatory, University of Warsaw, Al.~Ujazdowskie 4, 00-478 Warszawa, Poland}
\affiliation{Villanova University, Department of Astrophysics and Planetary Sciences, 800 Lancaster Ave., Villanova, PA 19085, USA}
\email{mwrona@astrouw.edu.pl}
\author{Mariusz Gromadzki}          
\affiliation{Astronomical Observatory, University of Warsaw, Al.~Ujazdowskie 4, 00-478 Warszawa, Poland}
\email{marg@astrouw.edu.pl}
\author{Mateusz J. Mr{\'o}z} 
\affiliation{Astronomical Observatory, University of Warsaw, Al.~Ujazdowskie 4, 00-478 Warszawa, Poland}
\email{mmroz@astrouw.edu.pl}
\collaboration{100}{(The OGLE Team)}
\author{David P. Bennett}
\affiliation{Code 667, NASA Goddard Space Flight Center, Greenbelt, MD 20771, USA}
\affiliation{Department of Astronomy, University of Maryland, College Park, MD 20742, USA}
\email{bennett.moa@gmail.com}
\author{Aparna Bhattacharya}
\affiliation{Code 667, NASA Goddard Space Flight Center, Greenbelt, MD 20771, USA}
\affiliation{Department of Astronomy, University of Maryland, College Park, MD 20742, USA}
\email{aparna.bhattacharya@nasa.gov}
\author{Kotaro Daimon}
\affiliation{Department of Earth and Space Science, Graduate School of Science, The University of Osaka, Toyonaka, Osaka 560-0043, Japan}
\email{daimon@iral.ess.sci.osaka-u.ac.jp}
\author{Ryusei Hamada}
\affiliation{Department of Earth and Space Science, Graduate School of Science, The University of Osaka, Toyonaka, Osaka 560-0043, Japan}
\email{hryusei@iral.ess.sci.osaka-u.ac.jp}
\author{Yuki Hirao}
\affiliation{Institute of Astronomy, Graduate School of Science, The University of Tokyo, 2-21-1 Osawa, Mitaka, Tokyo 181-0015, Japan}
\email{hirao@ioa.s.u-tokyo.ac.jp}
\author{Shuma Makida}
\affiliation{Department of Earth and Space Science, Graduate School of Science, The University of Osaka, Toyonaka, Osaka 560-0043, Japan}
\email{makida@iral.ess.sci.osaka-u.ac.jp}
\author{Stela Ishitani Silva}
\affiliation{Department of Physics, The Catholic University of America, Washington, DC 20064, USA}
\affiliation{Code 667, NASA Goddard Space Flight Center, Greenbelt, MD 20771, USA}
\email{ishitanisilva@cua.edu}
\author{Shota Miyazaki}
\affiliation{Institute of Space and Astronautical Science, Japan Aerospace Exploration Agency, 3-1-1 Yoshinodai, Chuo, Sagamihara, Kanagawa 252-5210, Japan}
\email{miyazaki@ir.isas.jaxa.jp}
\author{Tutumi Nagai}
\affiliation{Department of Earth and Space Science, Graduate School of Science, The University of Osaka, Toyonaka, Osaka 560-0043, Japan}
\email{nagai@iral.ess.sci.osaka-u.ac.jp}
\author{Seiya Nakayama}
\affiliation{Department of Earth and Space Science, Graduate School of Science, The University of Osaka, Toyonaka, Osaka 560-0043, Japan}
\email{nakayama@iral.ess.sci.osaka-u.ac.jp}
\author{Kansuke Nunota}
\affiliation{Department of Earth and Space Science, Graduate School of Science, The University Osaka, Toyonaka, Osaka 560-0043, Japan}
\email{unota@iral.ess.sci.osaka-u.ac.jp}
\author{Ryo Ogawa}
\affiliation{Department of Earth and Space Science, Graduate School of Science, The University Osaka, Toyonaka, Osaka 560-0043, Japan}
\email{rogawa@iral.ess.sci.osaka-u.ac.jp}
\author{Ryunosuke Oishi}
\affiliation{Department of Earth and Space Science, Graduate School of Science, The University Osaka, Toyonaka, Osaka 560-0043, Japan}
\email{oishi@iral.ess.sci.osaka-u.ac.jp}
\author{Hideaki Ose}
\affiliation{Department of Earth and Space Science, Graduate School of Science, The University Osaka, Toyonaka, Osaka 560-0043, Japan}
\email{ose@iral.ess.sci.osaka-u.ac.jp}
\author[0000-0001-5069-319X]{Nicholas J. Rattenbury}
\affiliation{Department of Physics, University of Auckland, Private Bag 92019, Auckland, New Zealand}
\email{n.rattenbury@auckland.ac.nz}
\author[0000-0002-1228-4122]{Yuki K. Satoh}
\affiliation{College of Science and Engineering, Kanto Gakuin University, Yokohama, Kanagawa 236-8501, Japan}
\email{yukisato@kanto-gakuin.ac.jp}
\author{Takahiro Sumi}
\affiliation{Department of Earth and Space Science, Graduate School of Science, The University Osaka, Toyonaka, Osaka 560-0043, Japan}
\email{sumi@ess.sci.osaka-u.ac.jp}
\author{Daisuke Suzuki}
\affiliation{Department of Earth and Space Science, Graduate School of Science, The University Osaka, Toyonaka, Osaka 560-0043, Japan}
\email{dsuzuki@ess.sci.osaka-u.ac.jp}
\author[0000-0002-6510-0681]{Motohide Tamura}
\affiliation{Astrobiology Center, 2-21-1 Osawa, Mitaka-shi, Tokyo 181-8588, Japan}
\affiliation{Department of Astronomy, University of Tokyo, 7-3-1 Hongo, Bunkyo-ku, Tokyo 113-0033, Japan}
\email{motohide.tamura@nao.ac.jp}
\author{Takuto Tamaoki}
\affiliation{Department of Earth and Space Science, Graduate School of Science, The University Osaka, Toyonaka, Osaka 560-0043, Japan}
\email{tamaoki@iral.ess.sci.osaka-u.ac.jp}
\author{Sean K. Terry}
\affiliation{Code 667, NASA Goddard Space Flight Center, Greenbelt, MD 20771, USA}
\affiliation{Department of Astronomy, University of Maryland, College Park, MD 20742, USA}
\email{skterry@umd.edu}
\author{Chihiro Ueda}
\affiliation{Department of Earth and Space Science, Graduate School of Science, The University Osaka, Toyonaka, Osaka 560-0043, Japan}
\email{ueda@iral.ess.sci.osaka-u.ac.jp}
\author{Aikaterini Vandorou}
\affiliation{Code 667, NASA Goddard Space Flight Center, Greenbelt, MD 20771, USA}
\affiliation{Department of Astronomy, University of Maryland, College Park, MD 20742, USA}
\email{aikaterini.vandorou@utas.edu.au}
\author{Hibiki Yama}
\affiliation{Department of Earth and Space Science, Graduate School of Science, The University Osaka, Toyonaka, Osaka 560-0043, Japan}
\email{yama@iral.ess.sci.osaka-u.ac.jp}
\collaboration{100}{(The PRIME Collaboration)}
\correspondingauthor{\texttt{cheongho@astroph.chungbuk.ac.kr}, \texttt{leecu@kasi.re.kr}}

\begin{abstract}
We present the analysis of the microlensing event KMT-2026-BLG-0083, which was 
independently detected by the KMTNet, OGLE, and PRIME surveys and also monitored 
by the DREAMS survey.  The combined data provide dense coverage of two distinct 
short-duration anomalies in the light curve that cannot be reproduced by a standard 
binary-lens model.  Independent analyses of the two anomalies indicate that each is 
produced by a planetary companion to the lens, motivating a triple-lens interpretation.  
The modeling yields two pairs of degenerate solutions arising from the well-known 
inner--outer degeneracy, with each pair exhibiting two local solutions depending on 
whether the source passes above or below the distant planetary companion.  The 
preferred model indicates two giant planets with mass ratios $q_2=(5.25\pm 0.08)\times 
10^{-3}$ and $q_3=(3.04\pm 0.24)\times10^{-3}$ orbiting a common host.  Bayesian analysis 
indicates that the host is an M-dwarf star with a mass of $0.48^{+0.36}_{-0.28}~M_\odot$, 
hosting two giant planets with masses of $2.65^{+1.96}_{-1.55}~M_{\rm J}$ and 
$1.54^{+1.14}_{-0.90}~M_{\rm J}$.  The projected planet--host separations are 
$10.6^{+1.7}_{-2.2}$~au and $1.7^{+0.3}_{-0.3}$~au, placing the inner planet near the
 host's snow line and the outer planet at a substantially larger separation.  Thus, 
KMT-2026-BLG-0083L becomes the seventh confirmed multiple-planet system discovered 
through gravitational microlensing, providing another example of a cold giant planetary 
system orbiting a subsolar-mass star.
\end{abstract}


\section{Introduction} \label{sec:one}

Gravitational microlensing has become one of the principal techniques for
discovering extrasolar planets because it probes a region of parameter
space that is largely inaccessible to other detection methods.
Unlike the transit and radial-velocity techniques, whose sensitivities
favor planets orbiting close to nearby stars, microlensing is most
sensitive to planets beyond the snow line, where giant planets are
expected to form. Because the lens need not be luminous, the technique
is capable of detecting planets orbiting hosts spanning a wide range of
masses, including M dwarfs, brown dwarfs, white dwarfs, and stellar
remnants, throughout both the Galactic disk and bulge
\citep{Mao1991, Gould1992,Gaudi2012}.

More than 300 exoplanets have now been discovered through gravitational
microlensing.\footnote{NASA Exoplanet Archive}
These discoveries have established microlensing as a powerful tool for
investigating the demographics of cold planetary systems. Statistical
analyses of microlensing planets have measured the occurrence rates of
planets beyond the snow line, revealed the planetary mass-ratio
function, and demonstrated that cold super-Earths are common on
Jupiter-like orbits around low-mass stars
\citep{Suzuki2016,Zang2025}. Together with transit and
radial-velocity surveys, these results are providing a more complete
picture of planet formation over a broad range of planetary masses,
orbital separations, and host-star properties.

Although isolated planets dominate the current microlensing sample,
systems containing multiple planets are of considerably greater
scientific value. Whereas a single detected planet provides only limited
information about an individual planetary system, the detection of
multiple planets directly constrains its architecture, including the
distribution of planetary masses, orbital separations, and relative
configuration. Such systems therefore provide powerful tests of theories
of planet formation, migration, and long-term dynamical evolution.
Because the Solar System itself contains multiple giant planets beyond
the snow line, discovering analogous systems around other stars is
particularly important for understanding whether the Solar System is
representative of planetary systems 
in general \citep{Gould2010}.

Despite this importance, multiple-planet systems remain exceptionally
rare in microlensing. Prior to this work, only six such systems had been
reported. Remarkably, although the sample is still very small, several
common characteristics have already emerged. All of the host stars have
subsolar masses, and nearly all of the detected companions are giant
planets orbiting beyond the snow line. Furthermore, four of the six
previously known systems contain two Jovian-mass planets.
Whether these characteristics reflect the intrinsic population of
planetary systems or merely current observational biases remains
unknown. In the standard core-accretion scenario, giant planets are
expected to form less efficiently around lower-mass stars because their
protoplanetary disks generally contain less solid material available for
building massive planetary cores \citep{Pollack1996, Ida2005, Laughlin2004}. 
Expanding the sample of multiple-planet systems is therefore essential for 
assessing whether the apparent preference for giant planets orbiting 
subsolar-mass hosts is statistically significant or simply a consequence 
of small-number statistics.

In this paper, we present the analysis of the microlensing event
KMT-2026-BLG-0083. The light curve exhibits two distinct short-duration
anomalies that cannot be simultaneously reproduced by a conventional
binary-lens model. We show that the anomalies are naturally explained by
two planetary companions orbiting a common M-dwarf host. Under this
interpretation, KMT-2026-BLG-0083 becomes the seventh confirmed
multiple-planet system discovered through gravitational microlensing,
extending the still limited sample of planetary systems beyond the snow
line.

\begin{figure*}[t]
\centering
\includegraphics[width=16.0cm]{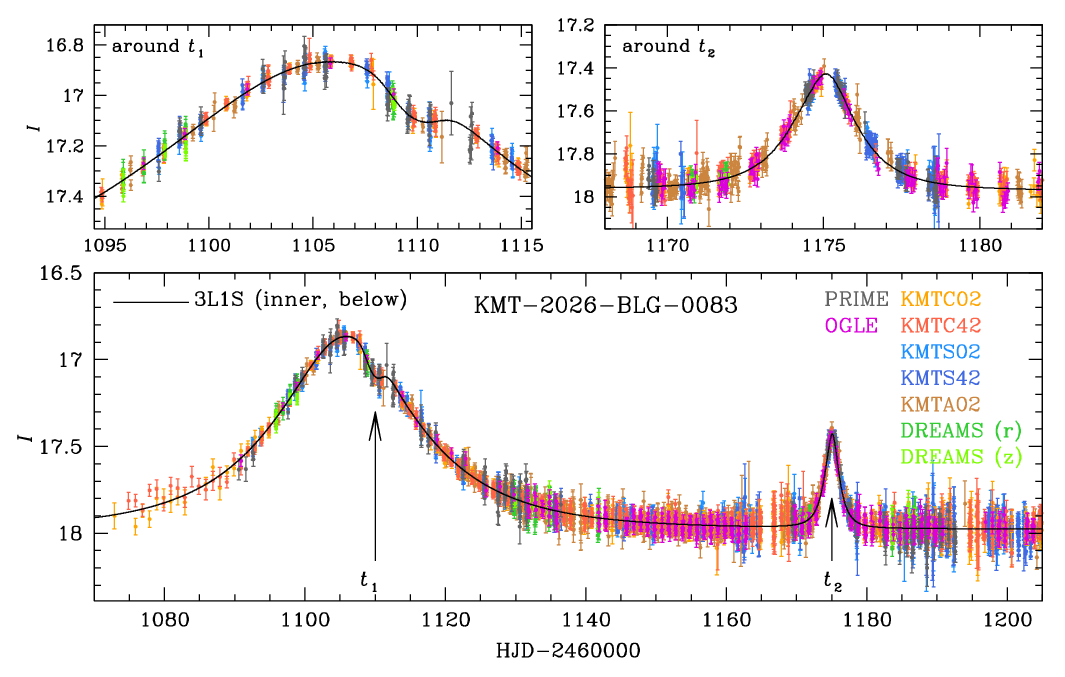}
\caption{
Light curve of the microlensing event KMT-2026-BLG-0083.  The lower panel presents the 
full light curve, while the upper two panels show enlarged views of the anomaly features 
indicated by the arrows labeled $t_1$ and $t_2$.  The solid curve overlaid on the data 
represents the best-fit 3L1S model. The colors of the data points correspond to those 
of the telescopes indicated in the legend.
}
\label{fig:one}
\end{figure*}

\section{Observation and data} \label{sec:two}

The microlensing event KMT-2026-BLG-0083 was discovered on 2026 March 16, corresponding 
to the abbreviated Heliocentric Julian Date ${\rm HJD}^\prime \equiv {\rm HJD}-2460000 =
1115$, by the Korea Microlensing Telescope Network (KMTNet; \citealt{Kim2016}) survey.  
The source is located at the equatorial coordinates $(\alpha, \delta)_{\rm J2000} = 
(17\!:\!53\!:\!36.31,\ \text{-}28\!:\!24\!:\!33.70)$, corresponding to the Galactic 
coordinates $(l, b) = (1\fdg 3521, \text{-}1\fdg 2317)$.  The event lies in the overlap region 
of the KMTNet prime fields BLG02 and BLG42.  Because each field was monitored with a 
cadence of 30 minutes, the effective cadence of the KMTNet observations was 15 minutes.  
The source has a baseline magnitude of $I_{\rm base}=18.18$, and the line of sight suffers 
an extinction of $A_I=2.74$.

The primary data set was obtained by the KMTNet survey, which operates three identical 
1.6\,m telescopes located at the Cerro Tololo Inter-American Observatory in Chile (KMTC), 
the South African Astronomical Observatory (KMTS), and the Siding Spring Observatory in 
Australia (KMTA).  Each telescope is equipped with a 4~deg$^2$ camera, allowing nearly 
continuous monitoring of the Galactic bulge through their longitudinal distribution.  
Most observations are obtained in the $I$ band, with occasional $V$-band exposures for 
source-color measurements.

Additional observations were provided by three independent surveys.  The Optical 
Gravitational Lensing Experiment (OGLE; \citealt{Udalski2015}) monitored the event with 
the 1.3\,m Warsaw Telescope at Las Campanas Observatory in Chile, obtaining observations 
primarily in the $I$ band with occasional $V$-band images.  The PRime-focus Infrared 
Microlensing Experiment (PRIME; \citealt{Sumi2025}) observed the event using the 1.8\,m 
telescope at the South African Astronomical Observatory, providing near-infrared photometry.  
The event was independently designated as OGLE-2026-BLG-0005 and PRIME-2026-BLG-0025 by 
the OGLE and PRIME surveys, respectively.
The event was also intensively monitored by the DECam Rogue Earths And Mars Survey 
(DREAMS; \citealt{Yang2026}), which employs the 4\,m Blanco Telescope equipped with the 
Dark Energy Camera (DECam) at CTIO.  DREAMS routinely observes the Galactic bulge with 
a cadence of 1--2 minutes, primarily in the $z$ band together with occasional $r$-band 
observations for color measurements.  Its large aperture and very high cadence provide 
dense coverage of short-timescale anomalies.

The KMTNet images were initially reduced using the difference-image-analysis (DIA) 
pipeline of \citet{Albrow2009} and were subsequently re-reduced using the pySIS package 
of \citet{Yang2024} to improve the photometric precision.  The OGLE photometry was obtained 
using the OGLE photometric pipeline \citep{Udalski2003}, the PRIME data were reduced using 
the PRIME photometric pipeline, and the DREAMS images were processed using the DECam 
photometric pipeline described by \citet{Yang2026}.

The photometric uncertainties of all data sets were renormalized following the prescription 
of \citet{Yee2012}, such that the cumulative $\chi^2$ distribution as a function of 
magnification is approximately linear and the reduced $\chi^2$ per degree of freedom is 
unity.  The final light curve used for the analysis combines the re-reduced KMTNet 
photometry with the OGLE, PRIME, and DREAMS observations.

\section{Light curve modeling} \label{sec:three}

Figure~\ref{fig:one} presents the observed light curve of KMT-2026-BLG-0083. Inspection of 
the light curve reveals two distinct anomaly features. The first anomaly occurs near the 
peak of the primary event, around ${\rm HJD}^\prime \sim 1110$ ($t_1$), and appears as a 
dip with a negative deviation from the underlying single-lens single-source model. 
The second anomaly is a short-lived bump with a positive deviation, occurring approximately 
65 days after the main event peak, around ${\rm HJD}^\prime \sim 1175$ ($t_2$). The dip 
anomaly lasts for approximately 3 days, whereas the bump persists for about 8 days.

We first investigate whether both anomaly features can be explained by a standard binary-lens 
single-source (2L1S) model. The light curve is characterized by seven lensing parameters. The 
first three, $(t_0, u_0, \te)$, denote the time of the closest lens--source approach, the 
lens--source separation (normalized to the angular Einstein radius, $\thetae$) at that time, 
and the event timescale, respectively.  The remaining four parameters describe the binary lens 
and the source trajectory: $(s, q)$ denote the projected separation (normalized to $\thetae$) 
and the mass ratio of the binary lens, $\alpha$ is the angle between the source trajectory and 
the binary-lens axis, and $\rho$ is the angular source radius normalized to $\thetae$. Despite 
an extensive search of the parameter space, we find no 2L1S solution that can simultaneously 
reproduce both observed anomaly features, indicating that a more complex lens or source 
configuration is required.

It is known that perturbations induced by multiple low-mass companions to the primary lens can
often be approximated as the superposition of the perturbations produced by the individual
companions \citep{Bozza2000,Han2001}. Under this superposition approximation, each anomaly can
be analyzed independently to identify the companion responsible for the corresponding
perturbation. Motivated by this, and by the inability of a single 2L1S model to reproduce both
anomalies simultaneously, we analyze each anomaly separately by excluding the data associated
with the other anomaly. This approach enables us to identify the origin of each perturbation
and provides suitable initial conditions for the subsequent full 3L1S modeling.

\begin{deluxetable*}{lll}
\tablewidth{0pt}
\tablecaption{2L1S solutions for central anomaly\label{table:one}}
\tablehead{
\multicolumn{1}{c}{Parameter}     &
\multicolumn{1}{c}{Inner    }     &
\multicolumn{1}{c}{Outer}
}
\startdata
$\chi^2$             &   6346.65                 &   6376.1                 \\
$t_0$ (HJD$^\prime$) &  $1105.663  \pm 0.016  $  &  $1105.750  \pm 0.017 $  \\
$u_0$                &  $    0.4434 \pm 0.0066$  &  $   0.4842 \pm 0.0091$  \\
$t_{\rm E}$ (days)   &  $   17.40   \pm 0.16  $  &  $  16.59   \pm 0.19  $  \\
$s  $                &  $    0.6642 \pm 0.0055$  &  $   0.9042 \pm 0.0082$  \\
$q  $ ($10^{-3}$)    &  $    3.14   \pm 0.23  $  &  $   3.38   \pm 0.29  $  \\
$\alpha$ (rad)       &  $    1.0843 \pm 0.0047$  &  $   1.1001 \pm 0.0051$  \\
$\rho$ ($10^{-3}$)   &   \nodata                 &   \nodata                \\
\enddata                                                                            
\tablecomments{HJD$^\prime = {\rm HJD} - 2460000$.}
\end{deluxetable*}

\subsection{Anomaly near the peak} \label{sec:three-one}

The dip feature around $t_1$ is likely to be of planetary origin because a short-duration 
negative deviation is a characteristic signature of a planetary perturbation. To test this 
interpretation, we modeled the light curve using a 2L1S configuration after excluding the 
data around $t_2$.  This analysis yielded a pair of solutions with binary-lens parameters 
$(s, q) \sim (0.67, 3.1 \times 10^{-3})$ and $(0.90, 3.4 \times 10^{-3})$, respectively.  
The low mass ratios confirm the planetary nature of the anomaly.  We refer to the solution 
as ``inner'' and ``outer'' solutions for the reason discussed below. The low mass ratio 
confirms the planetary nature of the perturbation. Both solutions have an event timescale 
of $\te \sim 17$~days.  The complete sets of lensing parameters for the two local solutions 
are listed in Table~\ref{table:one}. The inner solution is favored over the outer solution 
by $\Delta\chi^2 = 29.5$.

\begin{figure}[t]
\includegraphics[width=\columnwidth]{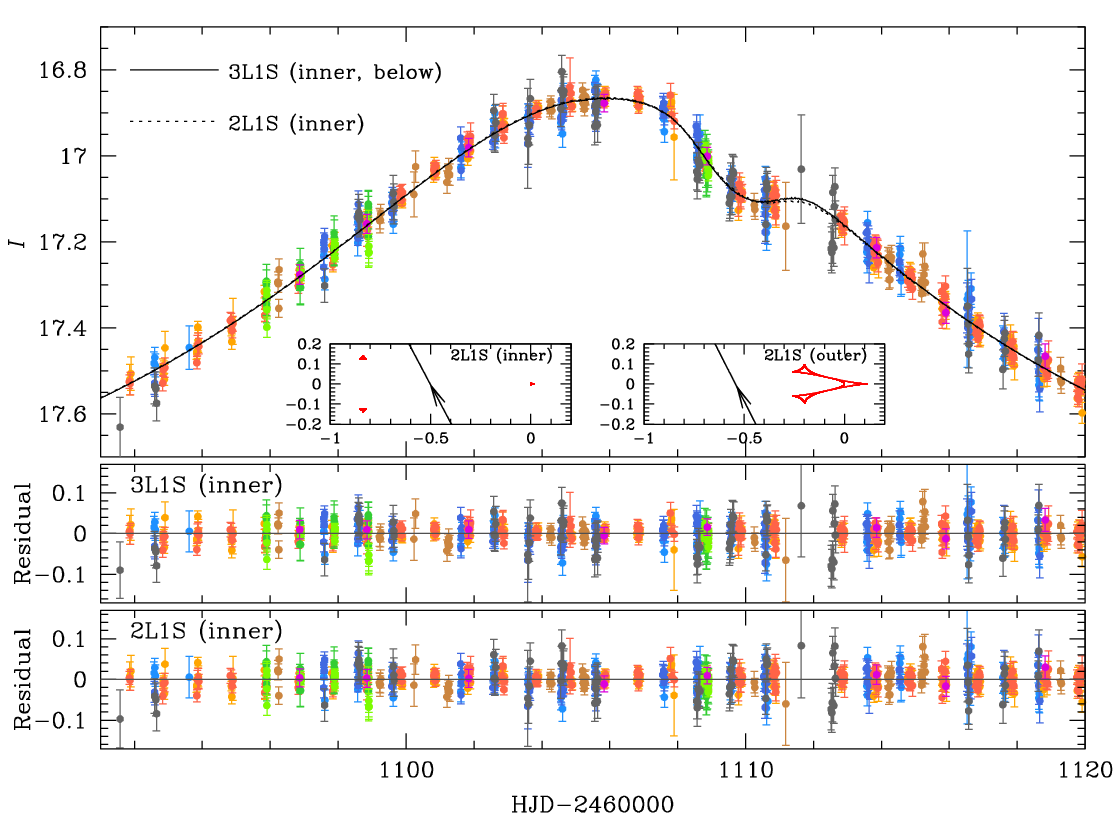}
\caption{
Enlarged view of the anomaly around $t_1$. The lower two panels show
the residuals from the best-fit 3L1S and 2L1S models. The corresponding model
curves are overlaid on the data, although they are indistinguishable within the line
width. The two insets in the upper panel present the lens configurations of the inner
and outer 2L1S solutions. In each inset, the red curves denote the caustics, and the
arrowed line represents the source trajectory.
}
\label{fig:two}
\end{figure}

Figure~\ref{fig:two} presents an enlarged view of the light curve around $t_1$, with the 
model curve of the inner 2L1S solution overlaid on the data.  The model reproduces the dip 
anomaly well, including both the negative deviation and the overall shape of the perturbation, 
supporting the interpretation that the feature is caused by a planetary companion.

The lens configurations of the two 2L1S solutions are shown in the insets of Figure~\ref{fig:two}. 
In both cases, the planetary companion induces two types of caustics: a central caustic centered 
on the primary lens and a pair of triangular planetary caustics located on the side of the host 
opposite the planet.  The two solutions differ only in the trajectory of the source
relative to the planetary caustics. In one solution, the source passes through the region between
the central and planetary caustics, whereas in the other it passes outside the planetary caustics.
We therefore designate these as the inner and outer solutions, respectively. Although the source
trajectories probe different regions of the magnification pattern, they produce nearly identical
perturbations, resulting in the well-known inner--outer degeneracy 
\citep{Gaudi1997, Yee2021, Gould2022, Zhang2022}.

\subsection{Anomaly in the Light Curve Wing} \label{sec:three-two}

In order to investigate the nature of the bump anomaly appearing in the wing of the light 
curve, we consider two possible scenarios for its origin: a low-mass companion to the lens 
and a faint companion to the source. The former is modeled using a 2L1S configuration, while 
the latter is modeled using a single-lens binary-source (1L2S) configuration. We additionally 
examine the 1L2S interpretation because it is known that a short-duration positive anomaly 
produced by a planetary companion can be mimicked by the close approach of a faint secondary 
source to the lens \citep{Gaudi1998,Gaudi2004}.

\begin{figure}[t]
\includegraphics[width=\columnwidth]{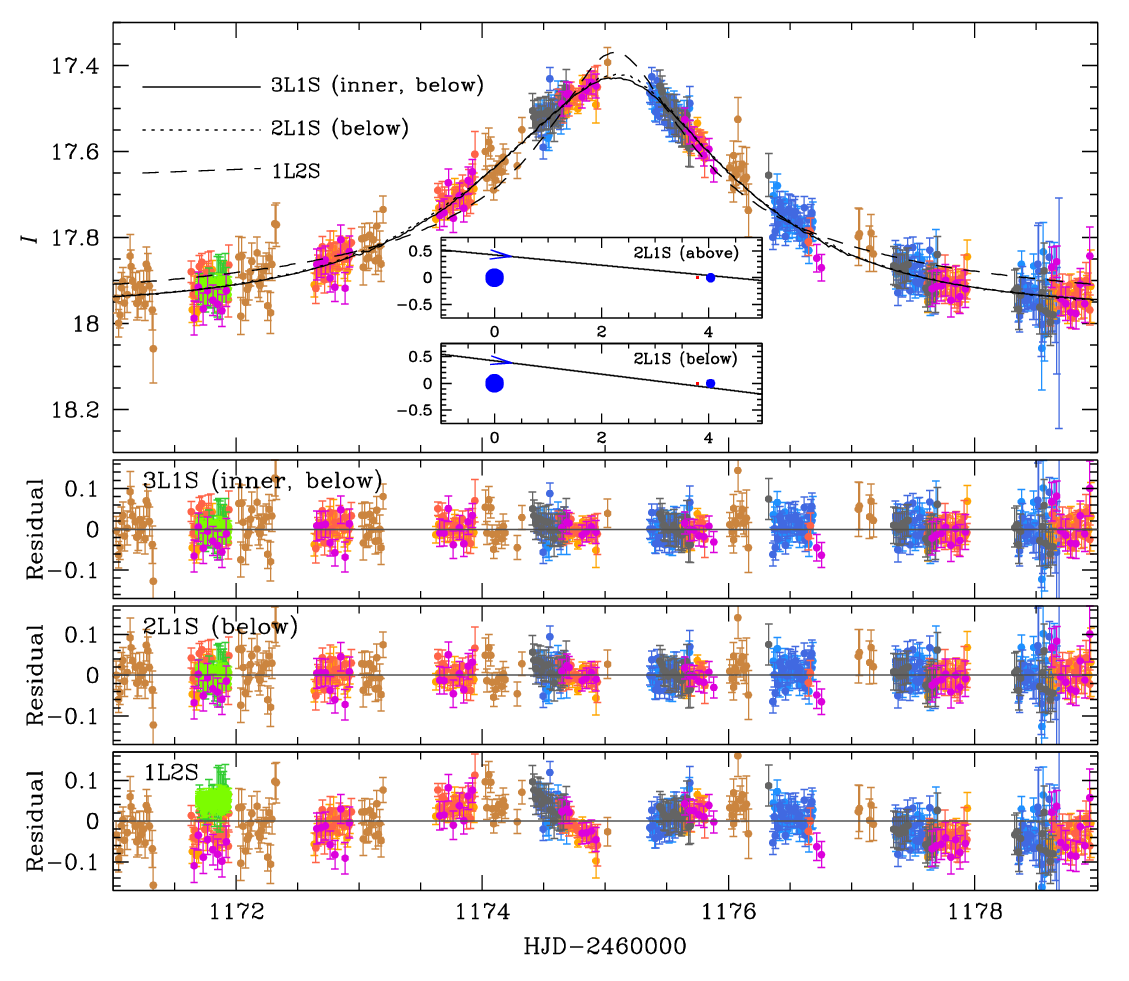}
\caption{
Close-up view of the anomaly around $t_2$. The 3L1S, 2L1S, and 1L2S model
curves are overlaid on the data. The lower panels show the residuals from the
corresponding models. 
The two insets in the upper panel present the lens configurations for the ``above'' and 
``below'' 2L1S solutions. In each inset, the blue dots mark the positions of the lens 
components, and the arrowed line represents the source trajectory.
}
\label{fig:three}
\end{figure}

\begin{deluxetable*}{lllllllll}
\tablewidth{0pt}
\tablecaption{Triple-lens solutions \label{table:two}}
\tablehead{
\multicolumn{1}{c}{Parameter} &
\multicolumn{2}{c}{Inner}     &
\multicolumn{2}{c}{Outer}     \\
\multicolumn{1}{c}{     }     &
\multicolumn{1}{c}{above}     &
\multicolumn{1}{c}{below}     &
\multicolumn{1}{c}{above}     &
\multicolumn{1}{c}{below}
}
\startdata
$\chi^2$             &    $8918.2                $  &  $8878.7                $  &  $ 8933.7                $  &  $8901.9               $  \\
$t_0$ (HJD$^\prime$) &    $1105.694   \pm 0.016  $  &  $1105.706   \pm 0.015  $  &  $ 1105.784   \pm 0.017  $  &  $1105.767  \pm 0.017  $  \\
$u_0$                &    $   0.4485  \pm 0.0066 $  &  $   0.4532  \pm 0.0064 $  &  $    0.4915  \pm 0.0075 $  &  $   0.4769 \pm 0.0069 $  \\
$t_{\rm E}$ (days)   &    $  17.23    \pm 0.15   $  &  $  17.11    \pm 0.15   $  &  $   16.34    \pm 0.15   $  &  $  16.65   \pm 0.14   $  \\
$s_2$                &    $   4.274   \pm 0.034  $  &  $   4.318   \pm 0.034  $  &  $    4.480   \pm 0.037  $  &  $   4.423  \pm 0.035  $  \\
$q_2$ ($10^{-3}$)    &    $   5.609   \pm 0.081  $  &  $   5.252   \pm 0.078  $  &  $    5.759   \pm 0.085  $  &  $   5.510  \pm 0.077  $  \\
$\alpha$ (rad)       &    $   3.23810 \pm 0.00072$  &  $   3.26645 \pm 0.00078$  &  $    3.24243 \pm 0.00071$  &  $   3.27034\pm 0.00078$  \\
$s_3$                &    $   0.6635  \pm 0.0054 $  &  $   0.6636  \pm 0.0055 $  &  $    0.8980  \pm 0.0074 $  &  $   0.8988 \pm 0.0076 $  \\
$q_3$ ($10^{-3}$)    &    $   3.06    \pm 0.23   $  &  $   3.04    \pm 0.24   $  &  $    3.36    \pm 0.27   $  &  $   3.16   \pm 0.28   $  \\
$\psi$ (rad)         &    $   4.1373  \pm 0.0051 $  &  $   4.1106  \pm 0.0051 $  &  $    4.1467  \pm 0.0046 $  &  $   4.1130 \pm 0.0046 $  \\
$\rho$               &    $    < 0.03            $  &  $   < 0.03             $  &  $    < 0.03             $  &  $   < 0.03            $  \\
\enddata                                                                            
\tablecomments{HJD$^\prime = {\rm HJD} - 2460000$.}
\end{deluxetable*}

Figure~\ref{fig:three} presents the model light curves and residuals of the best-fit 2L1S 
and 1L2S solutions.  Although both models reproduce the overall shape of the bump anomaly, 
the 2L1S model provides a substantially better fit to the data. The 1L2S model leaves 
systematic residuals throughout the anomaly, whereas these are largely removed by the 2L1S 
model. As a result, the 2L1S interpretation is favored over the 1L2S interpretation by 
$\Delta\chi^2 = 2561.1$, indicating that the bump anomaly is produced by a companion to the 
lens rather than to the source.

The 2L1S modeling yielded two local solutions.  The two insets in the upper panel of 
Figure~\ref{fig:three} show the corresponding lens-system configurations. In both solutions, 
the anomaly at $t_2$ is produced by a low-mass companion located at a projected separation exceeding 
four times the Einstein radius of the primary lens. In one solution, the source passes above 
the companion, whereas in the other it passes below it. We therefore refer to these as the 
``above'' and ``below'' solutions, respectively.  The ``below'' solution is slightly preferred, 
providing a better fit than the ``above'' solution by $\Delta\chi^2 = 9.0$.

The best-fit 2L1S model yields binary-lens parameters $(s, q) \sim (4.3, 5.3\times10^{-3})$.  
The low mass ratio indicates that the companion is of planetary mass, while the large projected 
separation implies that the planet is located at approximately four Einstein radii from its 
host.  Combined with the planetary interpretation of the first anomaly, this result strongly 
suggests that the two perturbations are produced by two distinct planetary companions orbiting 
the same host star.  This naturally motivates the full triple-lens single-source (3L1S) 
modeling presented in the next subsection.

\begin{figure*}[t]
\centering
\includegraphics[width=15.0cm]{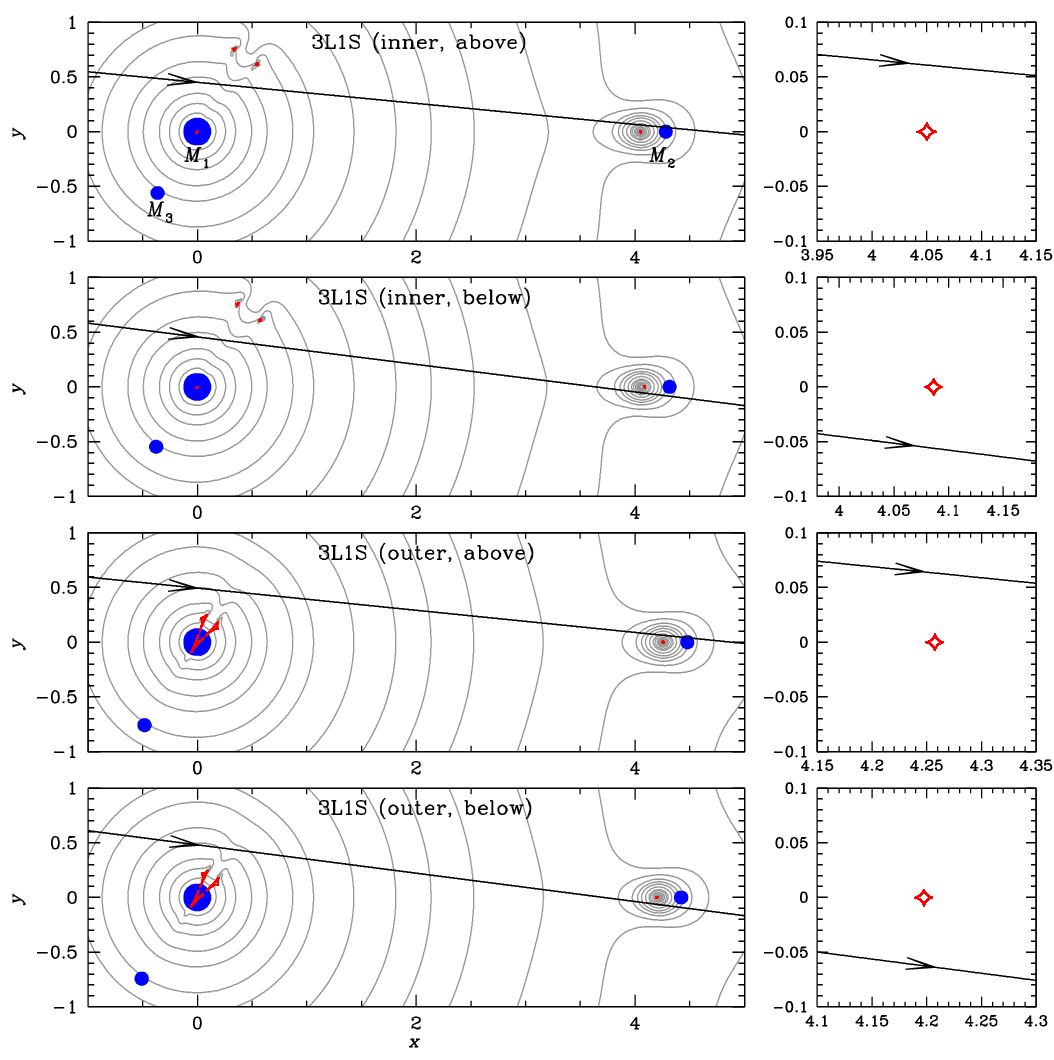}
\caption{
Lens configurations of the four local 3L1S solutions.  In each panel, the red curves denote 
the caustics, the arrowed lines represent the source trajectory, and the filled blue circles 
mark the positions of the lens components.  The gray curves surrounding the caustics are 
equi-magnification contours.  The panels on the right provide enlarged views of the planetary 
caustic induced by $M_2$, illustrating the source trajectory relative to the caustic.
}
\label{fig:four}
\end{figure*}

\subsection{Triple-lens model} \label{sec:three-three}

Having established that both anomaly features are of planetary origin, we proceed to model 
the full light curve using a 3L1S model. This model introduces three additional parameters 
to the standard 2L1S parameter set in order to describe the third lens component ($M_3$). 
These parameters are $(s_3, q_3, \psi)$, which represent the projected separation (normalized 
to $\thetae$) and mass ratio of $M_3$ relative to the primary lens ($M_1$), and the orientation 
angle of $M_3$ measured with respect to the $M_1$--$M_2$ axis.  Throughout this analysis, we 
assign $M_2$ and $M_3$ to the companions responsible for the anomalies around $t_2$ and $t_1$, 
respectively. Accordingly, we denote the projected separation and mass ratio of $M_2$ by 
$(s_2, q_2)$.

The 3L1S modeling is performed by carrying out a grid search over the parameters describing
$M_3$, namely $(s_3, q_3, \psi)$, while fixing the remaining parameters to the values obtained
from the 2L1S modeling of the $M_1$--$M_2$ binary pair. For each local minimum identified in
the grid search, we subsequently refine the solution by allowing all lensing parameters to vary
simultaneously. This procedure yields  four local solutions.

Table~\ref{table:two} lists the lensing parameters of the resulting 3L1S solutions. One pair 
of solutions is obtained by initiating the modeling from the inner 2L1S solution, while the 
other pair is obtained using the outer 2L1S solution as the initial guess. Within each pair, 
two local solutions arise depending on whether the source passes above or below the distant 
planetary companion. We therefore designate the four solutions as ``inner above,'' ``inner 
below,''  ``outer above,'' and ``outer below,'' respectively.  Among these, the ``inner below'' 
solution is preferred, yielding a lower $\chi^2$ than either of the other solutions by at least 
$\Delta\chi^2 = 23.2$.  The best-fit model curve for the preferred solution is shown in 
Figure~\ref{fig:one}, while enlarged views of the two anomaly regions are presented in 
Figures~\ref{fig:two} and \ref{fig:three}.

Consistent with the independent 2L1S analyses of the two anomaly features, both companions are
found to be of planetary mass, with mass ratios of $q_2 \sim 5.3 \times 10^{-3}$ and 
$q_3 \sim 3.0 \times 10^{-3}$.  The fractional uncertainty in $q_3$, $\sigma_{q_3}/q_3 \sim 7.5\%$, 
is substantially larger than that in $q_2$, $\sigma_{q_2}/q_2 \sim 1.5\%$.  This is expected because 
the anomaly around $t_1$, from which $q_3$ is determined, is much more sparsely sampled than the 
anomaly around $t_2$.

Figure~\ref{fig:four} presents the lens configurations of the four local 3L1S solutions. In all 
cases, each planetary companion generates its own planetary caustic. The companion $M_2$ produces 
a small four-cusp planetary caustic located near its position. The anomaly around $t_2$ is produced 
as the source passes close to this caustic, traversing its surrounding positive-magnification region. 
For each of the inner and outer solutions, the source can pass either above or below the companion, 
giving rise to the corresponding ``above'' and ``below'' solutions.  Although the source does not 
cross the caustic during this anomaly, its close approach places an upper limit on the normalized 
source radius of $\rho \lesssim 0.03$.

The second companion, $M_3$, is located off the $M_1$--$M_2$ axis at a position angle of 
$\psi \sim 236^\circ$ and induces a pair of triangular planetary caustics. In the inner 
solution, the source passes between the host and the planetary caustic, whereas in the outer 
solution it passes outside the planetary caustic. In both cases, however, the source traverses 
a negative-perturbation region adjacent to the caustic, producing the negative deviation 
observed around $t_1$.

\begin{figure}[t]
\includegraphics[width=\columnwidth]{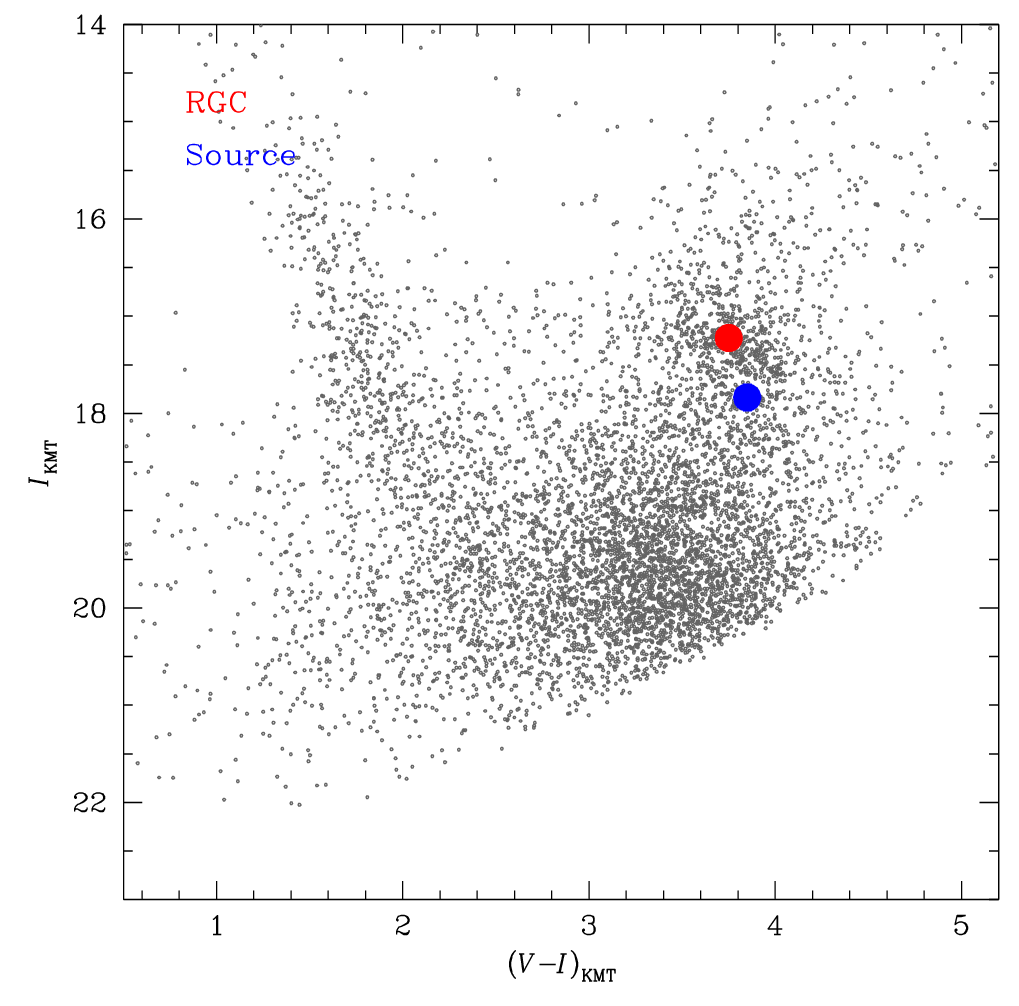}
\caption{
Instrumental color--magnitude diagram of stars in the vicinity of the source. The source 
and the centroid of the red giant clump (RGC) are marked. The RGC centroid is used to calibrate 
the source color and magnitude.
}
\label{fig:five}
\end{figure}

\section{Source star} \label{sec:four}

Characterizing the source star is an essential step in the analysis of a microlensing event
because it enables the estimation of the angular Einstein radius,
\begin{equation}
\theta_{\rm E} = \frac{\theta_*}{\rho},
\label{eq1}
\end{equation}
\hskip-4pt
from the angular source radius, $\theta_*$, and the normalized source radius, $\rho$.
The angular Einstein radius is related to the lens mass ($M$) and the distances to the
lens ($D_{\rm L}$) and source ($D_{\rm S}$) through
\begin{equation}
\theta_{\rm E} = \sqrt{\kappa M \pi_{\rm rel}}, \qquad
\pi_{\rm rel} = {\rm au}\left(\frac{1}{D_{\rm L}}-\frac{1}{D_{\rm S}}\right),
\label{eq2}
\end{equation}
\hskip-4pt
and therefore provides an important constraint on the physical properties of the lens.
Although the source does not cross a caustic during the event, the source passes 
sufficiently close to the planetary caustic induced by the more massive planet to place 
an upper limit on the normalized source radius, $\rho$. This, in turn, yields a lower 
limit on the angular Einstein radius. We therefore estimate the angular source radius 
by characterizing the source star.

To characterize the source, we first estimated its $V-I$ color by regressing the KMTC 
$V$- and $I$-band fluxes against the best-fit microlensing model. However, the resulting 
uncertainty was relatively large because the KMTNet $V$-band photometry had a low 
signal-to-noise ratio, owing to the high extinction toward the field and the modest event 
magnification. We therefore adopted an alternative estimate based on the DREAMS observations. 
Following the procedure described by \citet{Yang2026}, we first measured the source $r-z$ 
color by grouping temporally adjacent $r$- and $z$-band observations and performing a 
linear regression, obtaining $ r-z = 2.467 \pm 0.014$.  We then transformed this measurement 
to the KMTC photometric system using bright stars common to the DREAMS and KMTC catalogs in 
the vicinity of the event.

\begin{deluxetable}{lccclll}[t]
\tabletypesize{\small}
\tablewidth{0pt}
\tablecaption{Source parameters. \label{table:three}}
\tablehead{
\multicolumn{1}{c}{Parameter}           &
\multicolumn{1}{c}{Value}        
}
\startdata
 $(V-I)$                 &  \hspace{1cm}  $3.850 \pm 0.020 $   \hspace{1cm}  \\
 $I$                     &  \hspace{1cm}  $17.836 \pm 0.003$   \hspace{1cm}  \\
 $(V-I, I)_{\rm RGC}$    &  \hspace{1cm}  $(3.751, 17.225) $   \hspace{1cm}  \\
 $(V-I, I)_{\rm RGC,0}$  &  \hspace{1cm}  $(1.060, 14.388) $   \hspace{1cm}  \\
 $(V-I)_0$               &  \hspace{1cm}  $1.159 \pm 0.030 $   \hspace{1cm}  \\
 $I_0$                   &  \hspace{1cm}  $14.999 \pm 0.020$   \hspace{1cm}  \\
 Spectral type           &  \hspace{1cm}  K1 -- K2 III         \hspace{1cm}  \\
\enddata
\end{deluxetable}

The source position in the instrumental color--magnitude diagram is marked in 
Figure~\ref{fig:five}. Table~\ref{table:three} lists the instrumental and de-reddened 
color and magnitude of the source. The de-reddened values are determined using the 
method of \cite{Yoo2004}, in which the source is calibrated relative to the centroid 
of the red giant clump (RGC) in the same field. The table also lists the instrumental 
and de-reddened color and magnitude of the RGC centroid, $(V-I,\,I)_{\rm RGC}$ and 
$(V-I,\,I)_{{\rm RGC},0}$, respectively. Because the source and the RGC lie along 
nearly the same line of sight, they are expected to experience nearly the same extinction 
and reddening. The de-reddened source color and magnitude are then obtained by applying 
the measured offset of the source from the RGC centroid in the instrumental color--magnitude 
diagram to the known de-reddened RGC centroid \citep{Bensby2013,Nataf2013}.  The calibrated 
color and magnitude indicate that the source is a K1--K2 giant in the Galactic bulge.

The angular source radius was estimated from the calibrated source color and magnitude
using the standard color--surface brightness relation. We first converted the
de-reddened $(V-I)$ color to $(V-K)$ using the color--color relations of
\citet{Bessell1988}. The angular source radius was then derived from the
$(V-K)$--surface brightness relation of \citet{Kervella2004}, yielding
$\theta_* = 5.22 \pm 0.61~\mu{\rm as}$.
Combining this value with the upper limit on the normalized source radius,
$\rho \lesssim 0.03$, we obtained lower limits on the angular Einstein radius
and the relative lens--source proper motion,
\begin{equation}
\theta_{\rm E} > \theta_{\rm E,min} = \frac{\theta_*}{\rho_{\rm max}} = 0.17~{\rm mas},
\label{eq3}
\end{equation}
and
\begin{equation}
\mu > \mu_{\rm min} = \frac{\theta_{\rm E,min}}{t_{\rm E}} = 3.5~{\rm mas~yr^{-1}},
\label{eq3}
\end{equation}
respectively.

\begin{figure}[t]
\includegraphics[width=\columnwidth]{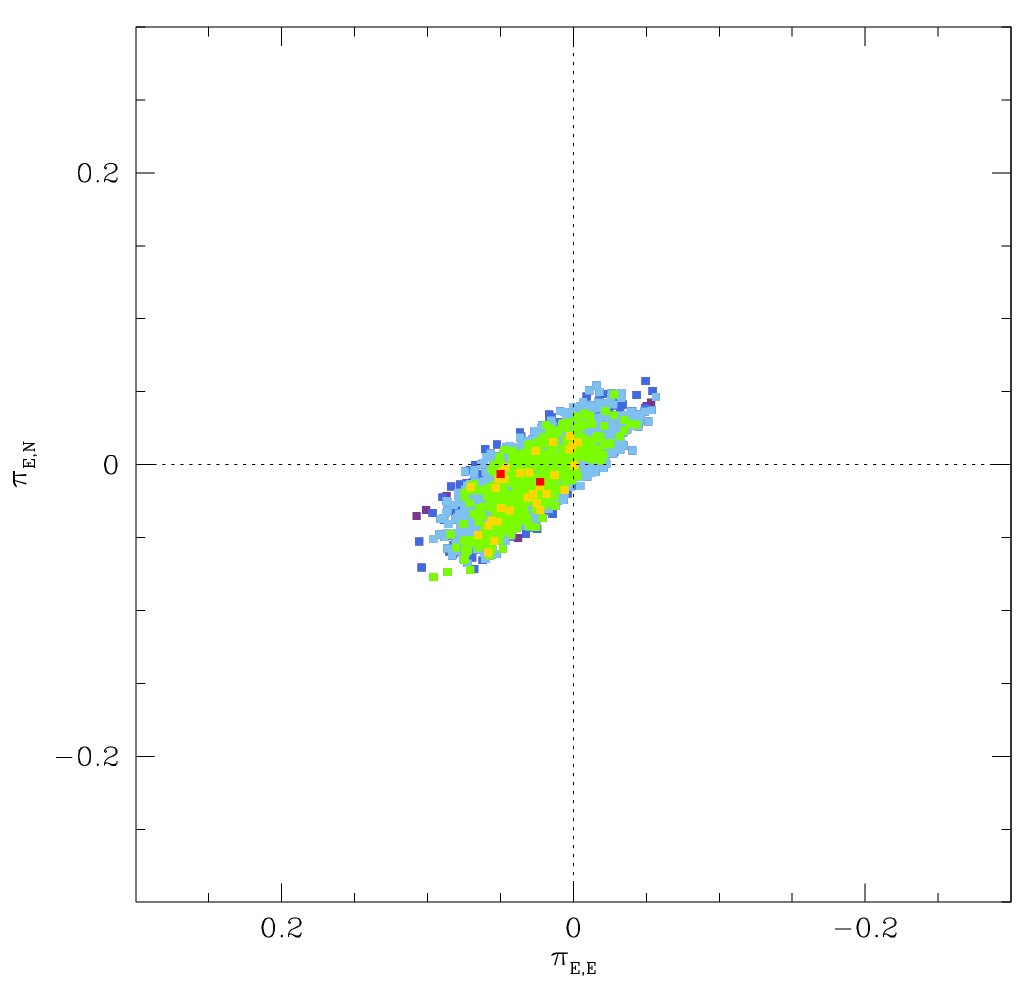}
\caption{
Distribution of the MCMC samples in the $(\piee,\pien)$ plane obtained from the parallax 
modeling.  The colors indicate the confidence levels: $1\sigma$ (red), $2\sigma$ (yellow), 
$3\sigma$ (green), $4\sigma$ (cyan), and $5\sigma$ (blue).
}
\label{fig:six}
\end{figure}

\section{Physical lens parameters} \label{sec:five}

The physical parameters of the lens, including its mass ($M$) and distance ($D_{\rm L}$), 
can be constrained from the lensing observables: the event timescale ($\te$), angular 
Einstein radius ($\thetae$), and microlens parallax ($\pie$). For KMT-2026-BLG-0083, the 
measured event timescale provides the primary observational constraint, while the lower 
limit on the angular Einstein radius provides an additional constraint. We therefore 
estimate the lens mass and distance through a Bayesian analysis.

\begin{figure}[t]
\includegraphics[width=\columnwidth]{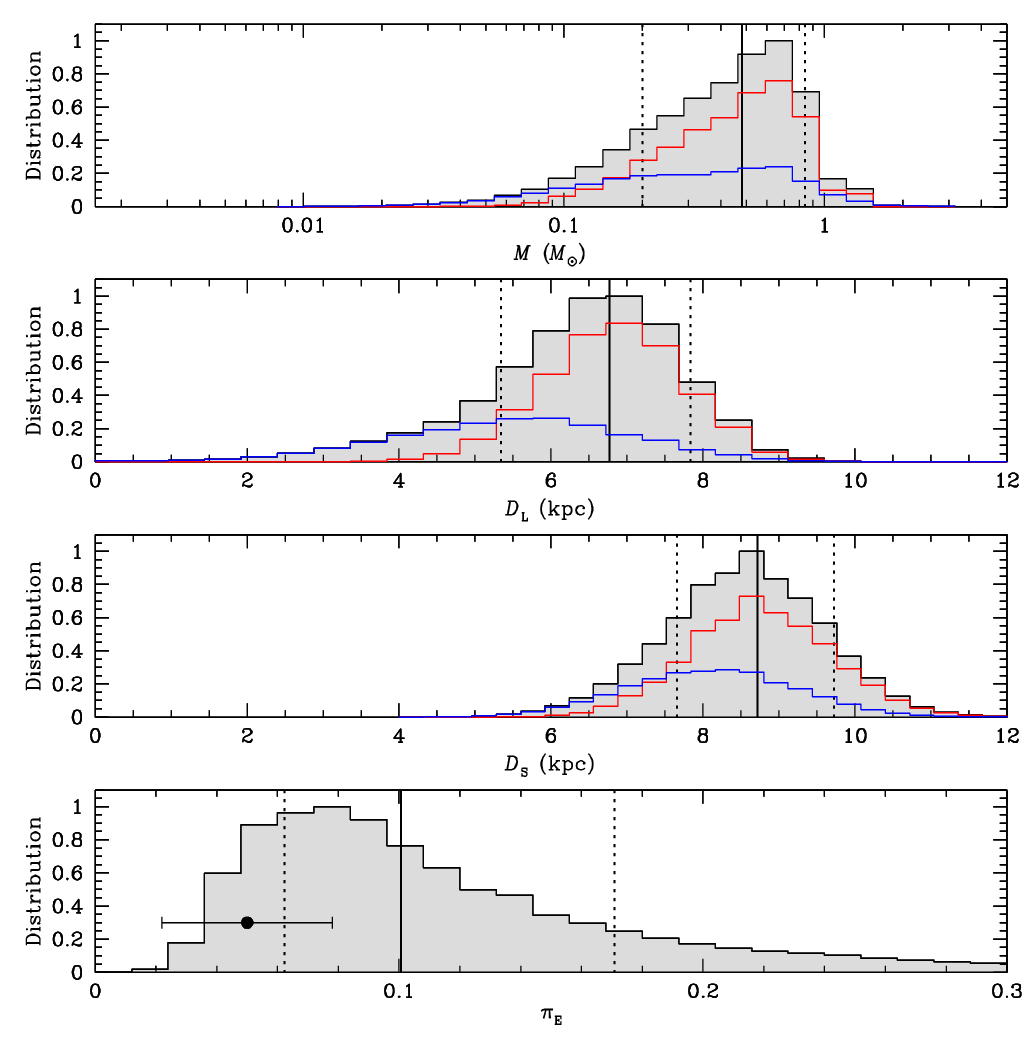}
\caption{
Bayesian posterior probability distributions of the primary lens mass, lens and source distances,
and microlens parallax. In the panels showing the distributions of $M$, $\dl$, and $\ds$,
the solid vertical line marks the median of each posterior distribution, while the two dotted
lines indicate the 1$\sigma$ confidence interval. The blue and red curves represent the
contributions from the Galactic disk and bulge lens populations, respectively, and the black
curve denotes their sum. In the panel showing the distribution of $\pie$, the point with an
error bar indicates the value and uncertainty of $\pie$ obtained from the parallax modeling.
}
\label{fig:seven}
\end{figure}

We also investigated whether the microlens-parallax effect could provide an additional 
constraint. Given the relatively short event timescale, $\te\sim17$~days, the parallax 
effect is expected to be weak. Nevertheless, modeling including the microlens-parallax 
parameters yields
\begin{equation}
(\pien, \piee)=(-0.007 \pm 0.023, 0.049 \pm 0.028),
\label{eq5}
\end{equation}
\hskip-4pt
with a negligible improvement in the fit.  Figure~\ref{fig:six} shows the distribution 
of samples from the Markov chain Monte Carlo (MCMC) analysis in the $(\pien,\piee)$ plane, 
illustrating that the parallax signal is not firmly detected.  The resulting constraint on 
$\pie$ is too weak to provide useful information on the physical lens parameters and is 
therefore not incorporated into the Bayesian analysis. We also find that the lensing 
parameters and their uncertainties obtained from the parallax model are nearly identical 
to those of the standard model. In particular, the inferred separation and its uncertainty 
for the wide-orbit planet remain essentially unchanged.

\begin{deluxetable*}{lccccll}
\tablewidth{0pt}
\tablecaption{Confirmed Multiple-planet Systems Discovered by Microlensing \label{table:four}}
\tablehead{
\colhead{System} &
\colhead{$M_{\rm h}$} &
\colhead{$M_{{\rm p},1}$} &
\colhead{$M_{{\rm p},2}$} &
\colhead{$a_{\perp,1}$} &
\colhead{$a_{\perp,2}$} &
\colhead{Reference} \\
\colhead{} &
\colhead{($M_\odot$)} &
\colhead{($M_{\rm J}$)} &
\colhead{($M_{\rm J}$)} &
\colhead{(au)} &
\colhead{(au)} &
\colhead{}
}
\startdata
OGLE-2006-BLG-109  & 0.51 & 0.71 & 0.27 & 2.3  & 4.6       & \citet{Gaudi2008, Bennett2010} \\
OGLE-2012-BLG-0026 & 0.82 & 0.11 & 0.68 & 3.8  & 4.6       & \citet{Han2013} \\
OGLE-2018-BLG-1011 & 0.18 & 0.15 & 0.27 & 1.8  & 0.8       & \citet{Han2019} \\
OGLE-2019-BLG-0468 & 0.92 & 3.4  & 10.2 & 3.3  & 2.8       & \citet{Han2022a} \\
KMT-2021-BLG-1077  & 0.14 & 0.22 & 0.25 & 1.3  & 0.9       & \citet{Han2022b} \\
KMT-2022-BLG-1818  & 0.78 & 2.6  & 0.65 & 2    & 0.4 / 13  & \citet{Li2026} \\
KMT-2026-BLG-0083  & 0.48 & 2.65 & 1.54 & 10.6 & 1.7       & This work \\
\enddata
\tablecomments{
$a_{\perp,1}$ and $a_{\perp,2}$ denote the projected separations of the first and 
second planets from their host star.}
\end{deluxetable*}

The Bayesian analysis combines the available observational constraints with a Galactic 
model and a lens mass function. The Galactic model specifies the spatial distribution 
and kinematics of the lens and source populations, while the lens mass function describes 
the distribution of lens masses. Based on these priors, a large ensemble of synthetic 
microlensing events is generated by assigning each realization a lens mass, lens and 
source distances, and a relative lens--source velocity drawn from the adopted distributions. 
The event timescale is then computed for each realization from the resulting physical 
parameters. In this work, we adopt the Galactic model of \citet{Jung2021} and the lens 
mass function of \citet{Jung2022}. By weighting the simulated events according to their 
consistency with the measured event timescale and imposing the lower limit on $\thetae$, 
we derive the posterior probability distributions of the lens mass and distance.

The posterior probability distributions of the host mass, lens and source distances 
are presented in Figure~\ref{fig:seven}.  The Bayesian analysis indicates that the 
lens system consists of two giant planets with masses
\begin{equation}
M_2 = 2.65^{+1.96}_{-1.55}~M_{\rm J},
\qquad
M_3 = 1.54^{+1.14}_{-0.90}~M_{\rm J}.
\label{eq6}
\end{equation}
\hskip-4pt
These planets orbit a host star with a mass
\begin{equation}
M_1 = 0.48^{+0.36}_{-0.28}~M_\odot,
\label{eq7}
\end{equation}
\hskip-4pt
corresponding to an M-dwarf star.

The lens system is estimated to be located at a distance of
\begin{equation}
\dl = 6.77^{+1.07}_{-1.43}~{\rm kpc},
\label{eq8}
\end{equation}
\hskip-4pt
while the source distance is estimated to be
\begin{equation}
\ds = 8.72^{+1.06}_{-1.00}~{\rm kpc}.
\label{eq9}
\end{equation}
\hskip-4pt
The combination of $\dl$ and $\ds$ yields a broad posterior distribution of 
$\pie={\rm au}(D_{\rm L}^{-1}-D_{\rm S}^{-1})$, peaking at approximately $\pie\sim0.075$, 
as shown in the bottom panel of Figure~\ref{fig:seven}. The point with an error bar marks 
the value and uncertainty of $\pie$ obtained from the parallax modeling. Although the 
parallax constraint is weak, the measured value is consistent with the Bayesian posterior 
distribution.

The Bayesian analysis indicates that the lens has a 34\% probability of belonging to the 
Galactic disk and a 66\% probability of belonging to the Galactic bulge.  The projected 
planet--host separations are
\begin{equation}
a_{\perp, 2} = 10.57^{+1.67}_{-2.23}~{\rm au}, \qquad
a_{\perp, 3} =  1.65^{+0.26}_{-0.34}~{\rm au}.
\label{eq10}
\end{equation}
\hskip-4pt
These values indicate that the inner planet ($M_3$) lies near the snow line of its host, 
while the outer planet ($M_2$) is located at a much larger projected separation.  We 
emphasize, however, that microlensing constrains only the projected planet--host separations. 
The true orbital separations may therefore be significantly larger than the values inferred 
here.

\section{Summary and Discussion} \label{sec:six}

We have presented the analysis of the microlensing event
KMT-2026-BLG-0083, whose light curve exhibits two distinct short-duration
anomalies. Independent analyses of the two perturbations show that each
is produced by a planetary companion to the lens, naturally motivating a
3L1S interpretation. The preferred 3L1S
solution indicates that the lens system consists of an M-dwarf host
orbited by two Jovian-mass planets. The inner planet lies near the
host's snow line, while the outer planet is located at a substantially
larger projected separation.

Table~\ref{table:four} summarizes the seven confirmed
multiple-planet systems discovered through gravitational microlensing,
including KMT-2026-BLG-0083L. Although the sample remains small, several
common characteristics are already apparent. All host stars have
subsolar masses, ranging from low-mass M dwarfs to nearly solar-mass
stars. In addition, nearly all detected companions are giant planets
orbiting beyond the snow line. KMT-2026-BLG-0083 follows these general
trends, further strengthening the emerging picture of cold giant
multiple-planet systems around low-mass stars.

Although the present sample is still too small to draw robust
statistical conclusions, the emerging demographic trends are
interesting. In the standard core-accretion scenario, the formation of
giant planets is expected to become less efficient around lower-mass
stars because their protoplanetary disks generally contain less solid
material available for building massive planetary cores \citep{Laughlin2004}. 
Whether the apparent prevalence of giant planets orbiting subsolar-mass hosts
reflects the intrinsic Galactic planet population or is primarily a
consequence of observational selection effects remains an open question.
Continued discoveries of multiple-planet systems will be essential for
distinguishing between these possibilities.

Future microlensing surveys, particularly the Nancy Grace Roman Space
Telescope \citep{Penny2019} and the Earth 2.0 (ET) mission \citep{Ge2022},
together with continued high-cadence ground-based surveys,
are expected to increase the number of known multiple-planet systems
substantially. The improved photometric precision and temporal coverage
of these surveys will enhance sensitivity to low-mass and widely separated
companions, thereby probing planetary architectures that are difficult to
access with other detection methods. A much larger and more uniformly
selected sample will enable statistical characterization of the frequency,
mass ratios, and orbital separations of planets in multiple-planet systems
beyond the snow line, and will provide stringent observational tests of
theories of planet formation and dynamical evolution.

\begin{acknowledgments}
C.H. was supported by the Chungbuk National University 2025 NUDP program and the National 
Research Foundation of Korea (RS-2025-21073000).
This research has made use of the KMTNet system operated by the Korea Astronomy and Space 
Science Institute (KASI) at three host sites of CTIO in Chile, SAAO in South Africa, and 
SSO in Australia. Data transfer from the host site to KASI was supported by the Korea 
Research Environment Open NETwork (KREONET). This research was supported by KASI under 
the R\&D program (project No. 2026-1-904-01) supervised by the Ministry of Science and ICT.
The OGLE project has received funding from the Polish National Science Centre grant OPUS-28
2024/55/B/ST9/00447 awarded to AU.

H.Y., W.Z. and S.M. acknowledge support by the National Natural Science Foundation of China 
(Grant No. 12133005, PI: S.M.). H.Y. acknowledge support by the China Postdoctoral Science 
Foundation (No. 2024M762938). The authors acknowledge the High-performance Computing Center 
at Westlake University for providing computational and data storage resources that have 
contributed to the research results reported within this paper. This work is part of the ET 
space mission, which is funded by the China's Space Origins Exploration Program. The work of 
K.B. is supported by NOIRLab, which is managed by the Association of Universities for Research 
in Astronomy (AURA) under a cooperative agreement with the U.S. National Science Foundation.
%
This project used data obtained with the Dark Energy Camera (DECam), which was constructed 
by the Dark Energy Survey (DES) collaboration. Funding for the DES Projects has been provided 
by the U.S. Department of Energy, the U.S. National Science Foundation, the Ministry of 
Science and Education of Spain, the Science and Technology Facilities Council of the United 
Kingdom, the Higher Education Funding Council for England, the National Center for Supercomputing 
Applications at the University of Illinois at Urbana-Champaign, the Kavli Institute for Cosmological 
Physics at the University of Chicago, the Center for Cosmology and Astro-Particle Physics at The 
Ohio State University, the Mitchell Institute for Fundamental Physics and Astronomy at Texas A\&M 
University, Financiadora de Estudos e Projetos, Funda\c{c}\~ao Carlos Chagas Filho de Amparo \`a 
Pesquisa do Estado do Rio de Janeiro, Conselho Nacional de Desenvolvimento Cient\'{\i}fico e 
Tecnol\'ogico and the Minist\'erio da Ci\^encia, Tecnologia e Inova\c{c}\~ao, the Deutsche 
Forschungsgemeinschaft, and the collaborating institutions in the Dark Energy Survey.
%
The collaborating institutions are Argonne National Laboratory; the University of California 
at Santa Cruz; the University of Cambridge; Centro de Investigaciones Energ\'eticas, 
Medioambientalesy Tecnol\'ogicas (CIEMAT), Madrid; the University of Chicago; University 
College London; the DES-Brazil Consortium; the University of Edinburgh; the Eidgen\"ossische 
Technische Hochschule (ETH) Z\"urich; Fermi National Accelerator Laboratory; the University 
of Illinois at Urbana-Champaign; the Institut de Ci\`encies de l'Espai (IEEC/CSIC); the 
Institut de F\'{\i}sica d'Altes Energies (IFAE); Lawrence Berkeley National Laboratory; the 
Ludwig-Maximilians-Universit\"at M\"unchen and the associated Excellence Cluster Universe; the 
University of Michigan; NSF NOIRLab; the University of Nottingham; The Ohio State University; 
the OzDES Membership Consortium; the University of Pennsylvania; the University of Portsmouth; 
SLAC National Accelerator Laboratory; Stanford University; the University of Sussex; and Texas 
A\&M University.
%
This project used data from the DECam Rogue Earths and Mars Survey (DREAMS), whose primary 
light-curve generation and archive are hosted by the Department of Astronomy at Westlake 
University. Based on observations at NSF Cerro Tololo Inter-American Observatory, NSF NOIRLab 
(NOIRLab Prop.\ ID 2025A-806294, PI: Weicheng Zang; Prop.\ ID 2025B-560332, PI: Weicheng Zang 
\& Hongjing Yang), which is managed by the Association of Universities for Research in Astronomy 
(AURA) under a cooperative agreement with the U.S. National Science Foundation.
\end{acknowledgments}



\bibliographystyle{aasjournal}
\bibliography{pasp_refs}

\end{document}